\documentstyle[epsfig,longtable]{aipproc}

\def\beq{\begin{equation}}
\def\eeq{\end{equation}}

\def\beqa{\begin{eqnarray}}
\def\eeqa{\end{eqnarray}}
\def\jpsi{J\!/\!\psi}

\begin{document}
\title{Hadronic Production of Heavy Quarks}
\footnotetext[1]{Review talk given at the b20 Symposium ``Twenty Beautiful
Years of Bottom Physics'', Illinois Institute of Technology, Chicago, 
June 29 -- July 2, 1997.}

\author{Matteo Cacciari$^2$}
\footnotetext[2]{Present address: LPTHE, Universit\'e Paris-Sud, 
Orsay, France}
\address{Deutsches Elektronen-Synchrotron DESY\\
D-22603 Hamburg, Germany}

\maketitle

\begin{abstract}
We review the status of theoretical evaluations of heavy quark and heavy
quarkonium hadroproduction cross sections and their comparisons with
experimental data.
\end{abstract}

\section*{Introduction}

When, more than twenty years ago, charm was discovered \cite{Aubert74} and 
subsequently interpreted \cite{Appelquist75} as the first heavy quark 
ever observed, it came as a 
big surprise. The discovery of bottom \cite{lederman} a few years later 
produced perhaps (I wasn't around at that time...) less excitement, but was 
surely equally important in extending our knowledge of what is now called the
heavy quark sector.

This sector is today well known, and probably recently completed by the
discovery of the top quark \cite{top}. Theoretical physics of heavy quarks has
therefore shifted gear, and moved from ``discovery mode'' to ``precision
physics.''

The new name of the game is now testing Quantum 
Chromodynamics, by checking its predictions against experimental results. The
latter have now grown quite accurate, and therefore demand equally precise
theoretical calculations. In this talk I shall
describe the state of the art of such calculations for heavy quark 
hadroproduction. 

I shall first review the fixed-order next-to-leading order (NLO) QCD
calculation, now  available for total cross sections and one-  and
two-particles distributions. It is a consolidated result and provides a
benchmark for future developments.

Large logarithms  appear in this calculation and
potentially make it less reliable in some regimes: $\log(S/m^2)$ 
and $\log(p_T^2/m^2)$ become large when the center of mass energy $\sqrt{S}$ or
the transverse  momentum $p_T$ of the
observed quark is much larger than its mass. Large $\log(1-4m^2/\hat s)$ appear when
the heavy quarks are produced close to the partonic threshold.
I shall briefly describe the resummation of $\log(1-4m^2/\hat s)$ and 
$\log(p_T^2/m^2)$ terms, and the inclusion of non-perturbative fragmentation
effects.

I shall finally also briefly comment on the subject of heavy quarkonium
production, where our understanding seems to have been greatly
increased by a lot of recent theoretical activity.
 
\section{NLO calculation}

The road to the NLO evaluation of heavy quark hadroproduction cross sections
was paved by Collins, Soper and Sterman \cite{css}, who argued that the 
following factorization formula holds:
\beq
d\sigma(H_1 H_2 \to Q\overline{Q}; m) = 
\sum_{ij} \int f_{i/H_1} f_{j/H_2} d\hat\sigma(ij\to Q\overline{Q}; m) + {\cal
O}\left({\Lambda_{QCD}\over m}\right) \; ,
\label{QQfact}
\eeq
where the summation over partons $i$ and $j$ runs only over gluons and 
light quarks,
and the heavy quarks are generated only at the perturbative level, by gluon
splitting. The cross section explicitly depends on the heavy quark mass $m$ and
on all other scales entering the problem (total energy $\sqrt{S}$, transverse
momentum $p_T$, etc.).

Along these lines, explicit calculations were performed by two groups, 
Nason, Dawson and Ellis \cite{NDE} on one side and Beenakker, Kuijf, Meng, van
Neerven, Schuler and Smith \cite{beenakker}
on the other. More recently Mangano, Nason, and Ridolfi (MNR) \cite{MNR} have
presented a Montecarlo integrator, based on the first of the two
calculations, which can provide fully exclusive cross sections, 
thereby allowing detailed
comparisons with experimental data. A very extensive collection of such
comparisons is presented in a recent review\cite{fmnr-rev}, from which we
select some plots to be shown here.

\begin{figure}[t]
\begin{center}
\begin{minipage}{6.5truecm}
\epsfig{file= bcproton_96.eps,width=6.5truecm}
\end{minipage}
\begin{minipage}{6.5truecm}
\epsfig{file= 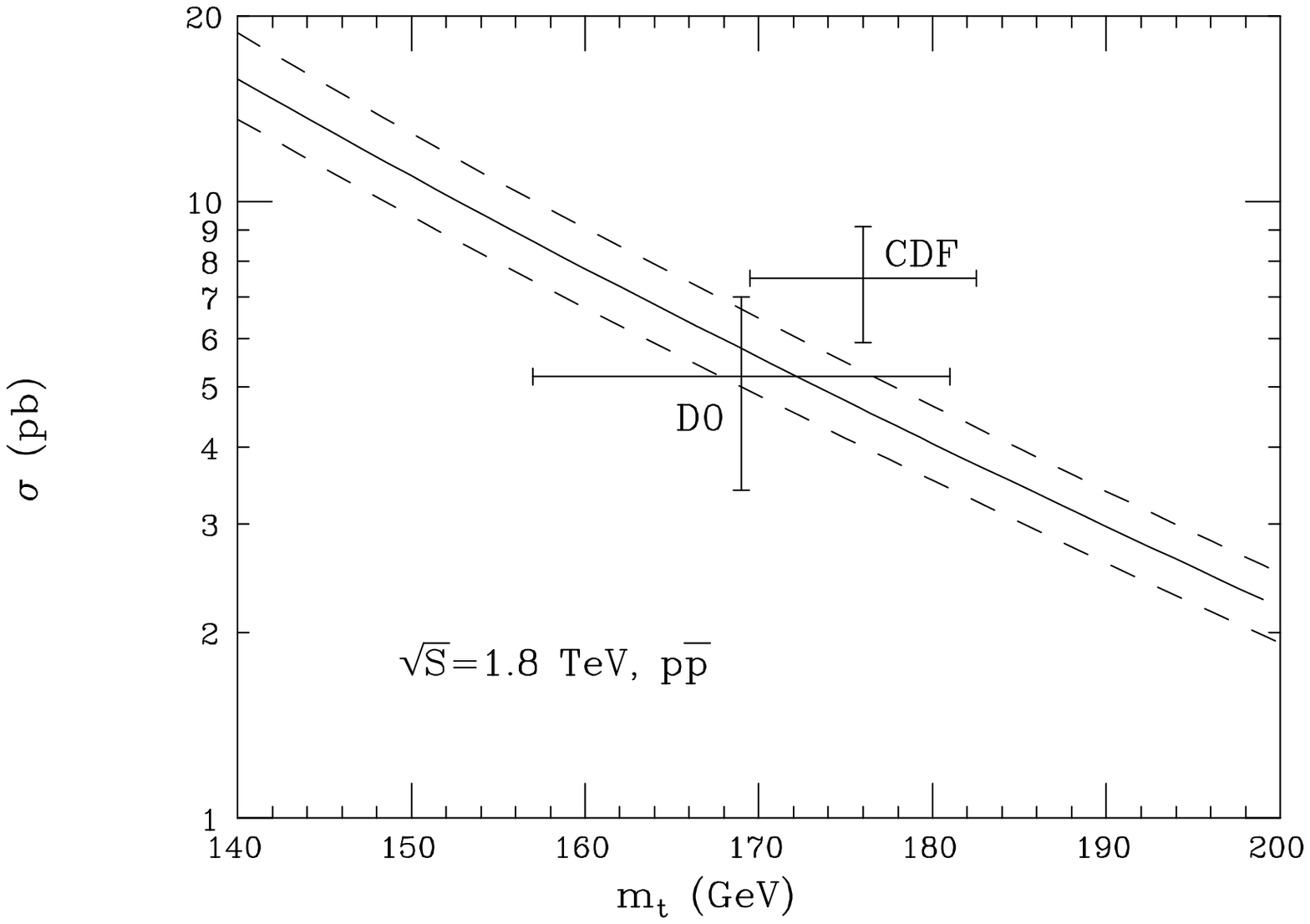,width=6.5truecm}
\end{minipage}
\end{center}
\caption{Charm and bottom production at fixed target experiments vs. NLO 
theoretical predictions (left), and top production at the Tevatron (right),
from ref. \protect\cite{fmnr-rev}.}
\label{fig1}
\end{figure}

The ratios of next-to-leading over leading order predictions for total cross 
sections depend on $m/\sqrt{S}$ and are about 1.3 for
top production at Tevatron energy ($\sqrt{S} = 1800$ GeV) and of order 2 or 
larger for charm and bottom already at fixed target energy.
Large uncertainties, due to monotonic renormalization/factorization 
scale dependence, are present for charm and 
bottom, while the prediction is fairly reliable for top ($\pm
10\%$), as shown in figure \ref{fig1}.

\begin{figure}[t]
\begin{center}
\begin{minipage}{6.5truecm}
\epsfig{file= 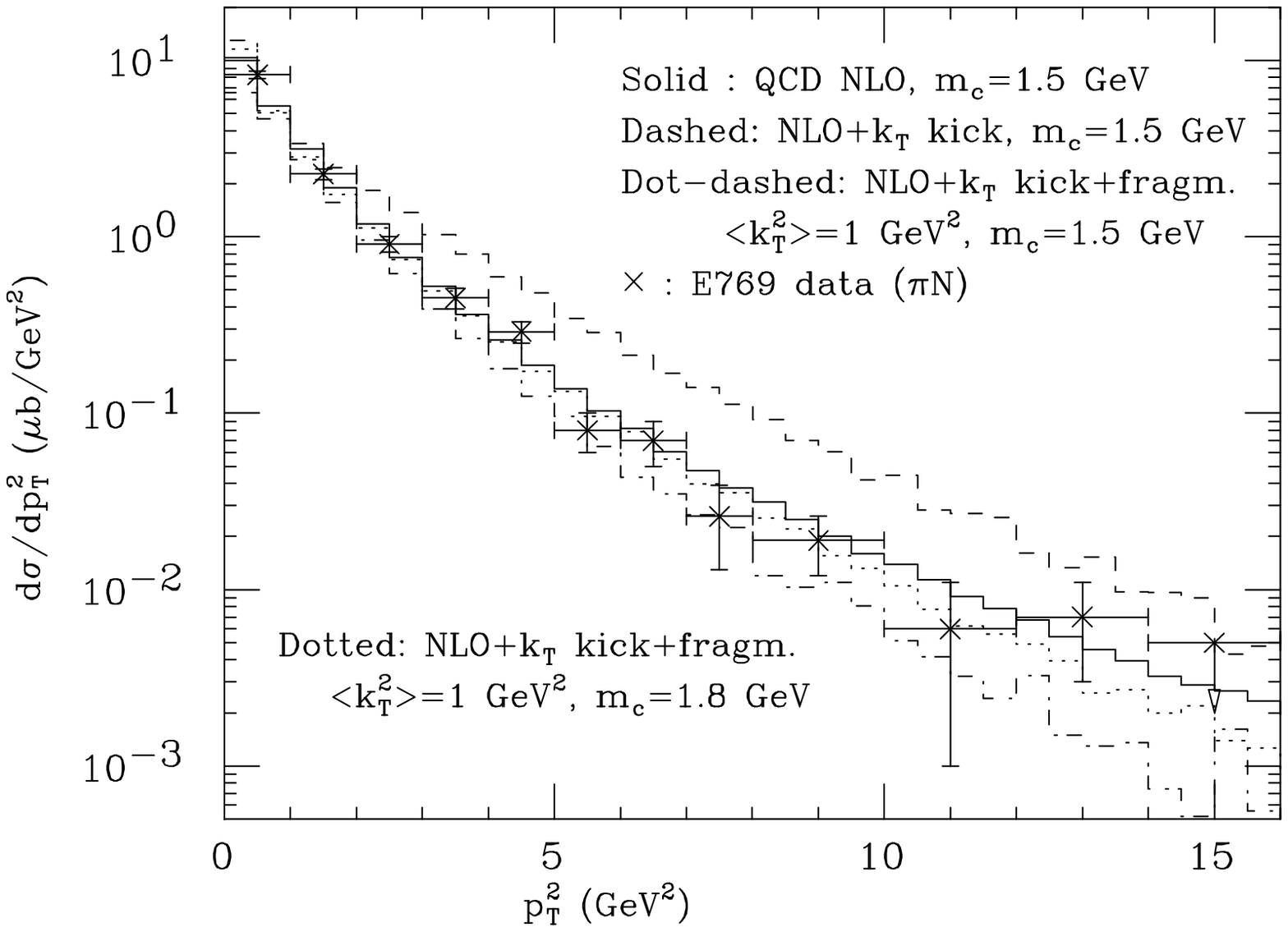,width=6.5truecm}
\end{minipage}
\begin{minipage}{6.5truecm}
\epsfig{file= 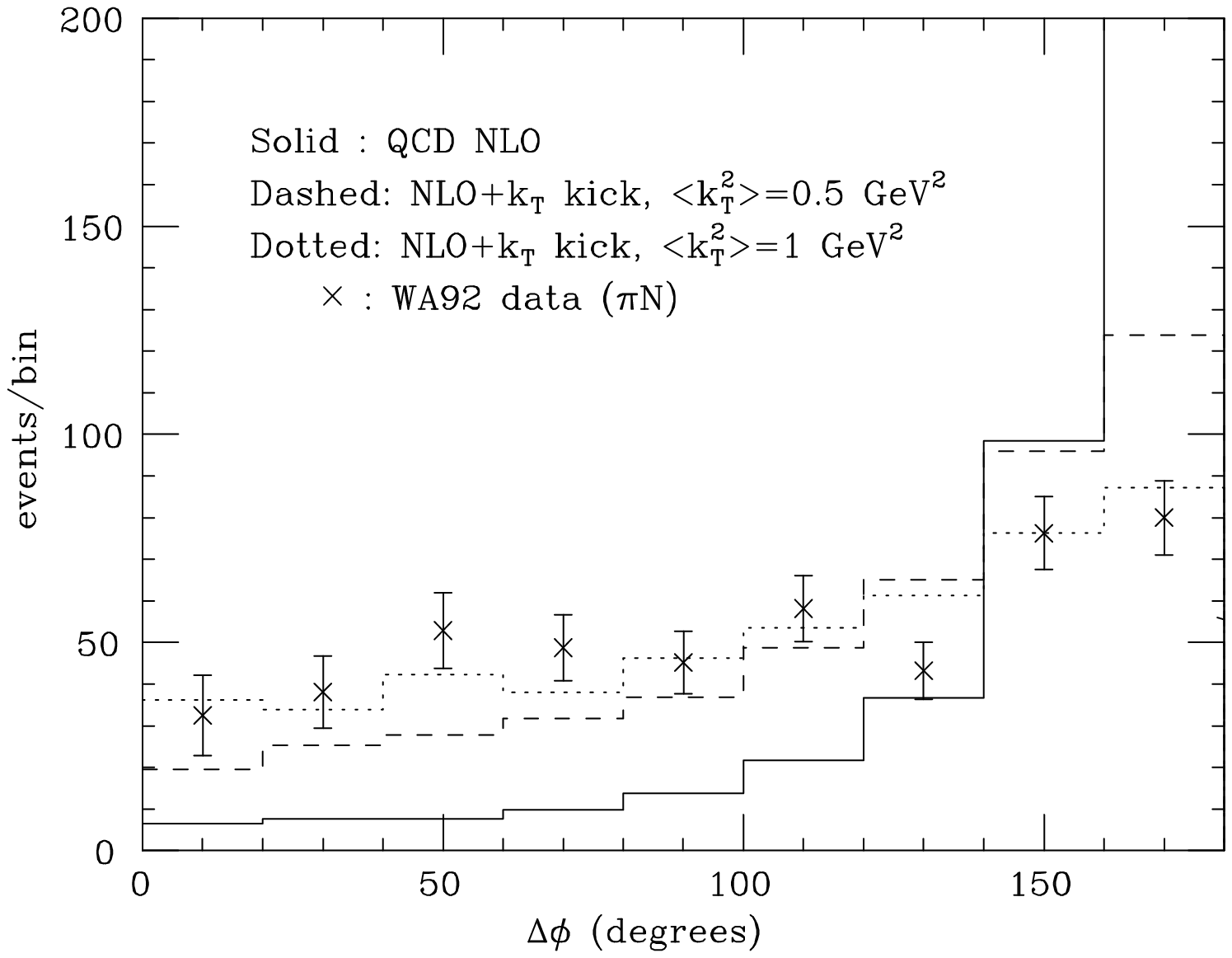,width=6.5truecm}
\end{minipage}
\end{center}
\caption{Charm $p_T$ inclusive distribution in $\pi N$ collisions vs.
theoretical predictions [NLO
calculation + initial $k_T$ ($\langle k_T^2\rangle = 1$ GeV$^2$) + 
non-pert.\ fragmentation (Peterson, $\epsilon = 0.06$)] on the left, and
azimuthal correlation of the $c\bar c$ pair on the right. 
From ref.~\protect\cite{fmnr-rev}.
}
\label{fig2}
\end{figure}

One-particle inclusive differential cross sections, usually $p_T$
distributions, can also be considered. Fig.~\ref{fig2} shows on the left 
a comparison of
E769 pion-nucleon fixed target data with NLO QCD plus two non-perturbative
contributions: an intrinsic initial transverse momentum $k_T$ of the colliding
partons, with $\langle k_T\rangle = 1$ GeV, and fragmentation effects of the
charm into the observed charmed hadrons, described with the aid of a Peterson
\cite{pssz}
fragmentation function (FF) with $\epsilon = 0.06$. The overall result
(the dot-dashed line) can be seen to be slightly softer than the
data. Uncertainties are however large: a larger mass, a larger $\langle
k_T\rangle$ or a harder FF could help improve the agreement. We should
mention that, while the inclusion of the non-perturbative contributions might
appear unnecessary here, it looks however mandatory when considering
two-particle distributions or even $p_T$ distributions in
photon-hadron collisions, pure NLO QCD being there clearly unable to describe
the data by itself. An example is given in fig.~\ref{fig2}, on the right,  
where WA92
pion-nucleon data for the
azimuthal distance $\Delta\phi$ of the $c$ and the $\bar c$ are compared to
theoretical predictions: the inclusion of the $k_T$ kick brings the curve into
agreement with the data. More such comparisons can be found in \cite{fmnr-rev}.
The overall picture suggests some consistency between the non-perturbative
inputs needed to describe the one- and the two-particle distributions
in both photo- and hadroproduction.

\begin{figure}[t]
\begin{center}
\begin{minipage}{6.5truecm}
\epsfig{file= 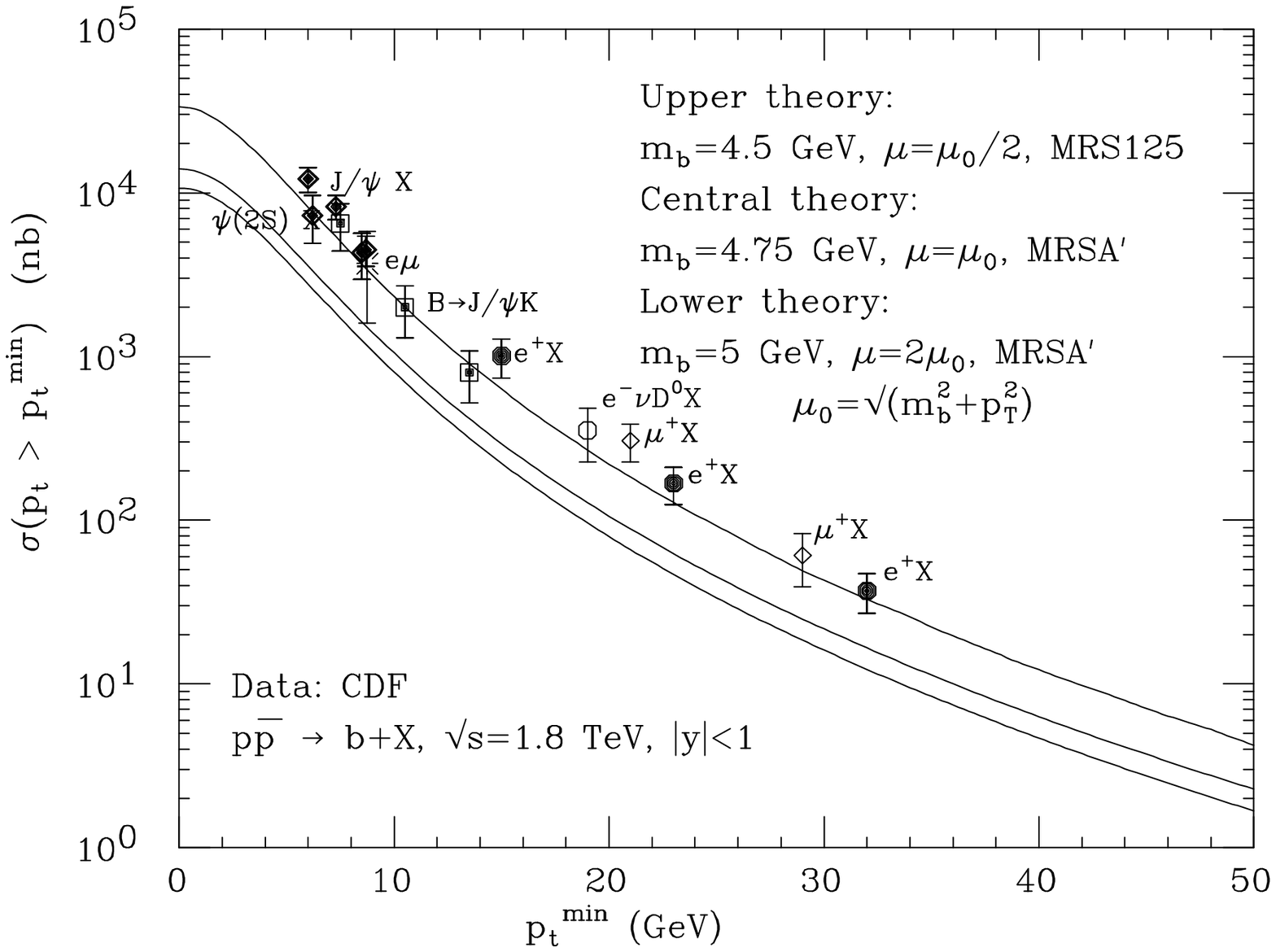,width=6.5truecm,height=5.5truecm}
\end{minipage}
\begin{minipage}{6.5truecm}
\epsfig{file= 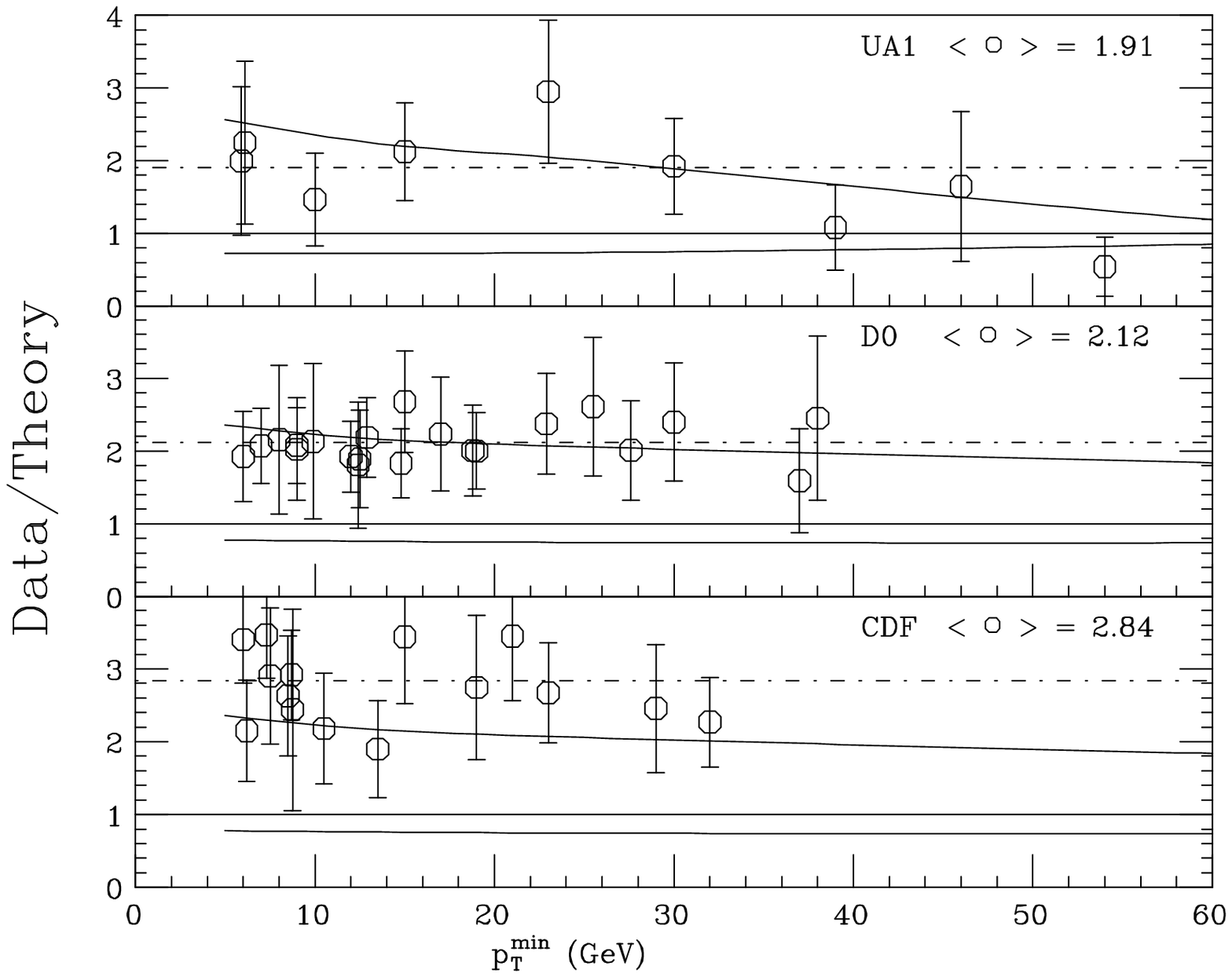,width=6.5truecm}
\end{minipage}
\end{center}
\caption{Bottom $p_T$ distributions data versus theoretical predictions,
from ref. \protect\cite{fmnr-rev}.}
\label{fig3}
\end{figure}

More comparisons of $p_T$ distributions can be done with the data for bottom
production taken at the Tevatron  by the CDF and D0 
experiments. These distributions initially caused some concern, as they were
markedly higher than the NLO QCD predictions. The data have now come down a
bit, thanks to a better understanding of some decay chains used in the analysis
(like the $B\to J\!/\!\psi\to \mu^+\mu^-$ one) and to the use of
microvertex silicon detectors which allow a much better identification and
reconstruction of the heavy quark events. But, still, theory and data are not
in perfect agreement. Fig.~\ref{fig3} shows on the left the data from CDF and
on the right the data/theory ratios from CDF, D0 and UA1. Data can be seen to
overshoot the central theoretical predictions by factors between two and three,
and to be in fair agreement with the upper edge of the theoretical uncertainty
band. One can thus conclude that no serious disagreement is present, but that
the situation certainly deserves further investigations, also in the light of 
data on forward bottom production by the D0 Collaboration, which appear to be
about a factor of four higher than the theoretical prediction.

\section{Resummation of threshold effects}

As we mentioned in the Introduction, potentially 
large logarithms  appear in the  NLO fixed-order calculation and
make it less reliable in some regimes. We deal now with $\log(1-4m^2/\hat s)$
terms, which appear in the partonic cross section in the form 
\beq
\hat\sigma(\hat s) = \sigma_0(\hat s)\left[1 + C \alpha_s {
\log^2(1-\frac{4m^2}{\hat s})} +
\cdots
\right] \; .
\eeq
These logarithms become large when the cms energy the colliding partons have is
close to the invariant mass of the two produced heavy quarks, that is to say,
when the quarks are produced close to the threshold. Resummation of such terms
can turn out to be phenomenologically important when the quark is very heavy
with respect to the hadronic collision energy: top production at the Tevatron,
with the top weighing 175 GeV and the machine delivering 1800 GeV, could be
such a case.

In order to better investigate this issue three groups have in
recent years attempted such a resummation. We are not going to go into any
details here, to be found in the respective papers, but simply summarize their
conclusions and point to similarities and differences of approaches and
results.

Laenen, Smith and van Neerven \cite{lsvn} performed the resummation in 
$x$-space, by
directly exponentiating the leading log (LL) $\alpha_s\log^2(1-x)$ term, where
$x=4m^2/\hat s$. Their formulae necessitate the introduction of an
artificial infrared cutoff which serves a double purpose: it avoids hitting 
the Landau pole when the now $z$-dependent argument of the running coupling 
$\alpha_s$ 
tends to zero as $z\to 1$, and it
regulates an otherwise divergent integral of the form $\int_0^1 dz
\exp[|a|\log^2(1-z)]$. They predict a
10\% increase over the NLO fixed order calculation for top production at the
Tevatron, but with a large uncertainty due to the dependence on the unphysical
cutoff.

Berger and Contopanagos \cite{bc} instead perform the resummation in 
Mellin moments
space, and then invert back to $x$-space for producing
phenomenological predictions. When performing this inversion they apply a
so-called principal value prescription to deform the integration contour and
avoid hitting the Landau pole. They then discard all the non-leading
terms generated by the inverse transform, arguing they are not universal, 
and only retain the LL ones. The further need
for an infrared cutoff to avoid the $\int_0^1 dz
\exp[|a|\log^2(1-z)]$ divergence is met by choosing it in such a way that
all integrations are confined to a perturbative domain, i.e. to regions where
the discarded non-leading terms are really subdominant. Through this procedure 
they also find an
increase of about 10\%, but claim a small uncertainty due to the motivated
choice of the cutoff.

A third evaluation of soft-gluon resummation effects has been performed by
Catani, Mangano, Nason and Trentadue \cite{cmnt}. They also perform 
the resummation in
Mellin space, and avoid the Landau pole by what they call minimal
prescription, i.e.\ a choice of the integration contour which leaves the
non-perturbative branch cut to the right. After doing so, they perform
the $z$ integration by retaining both the leading and the next-to-leading (NLL)
terms. Their argument is that the NLL contributions are generated by momentum
conservation and, while formally subleading, discarding them leads to
factorially divergent integrals and hence to the need for a cutoff. The
phenomenological outcome of their investigation, at variance with previous
findings, is that top hadroproduction at the Tevatron energy is predicted to
increase by only 1\% with respect to the fixed order result. They also claim
little uncertainty on this result due to the absence of an explicit cutoff.

Whatever the approach one considers correct, we are however still far from
experimentally probing such fine details (see experimental errors in
fig.~\ref{fig1}).

\section{Large Transverse Momentum Resummation}
Another kind of large logarithmic terms appearing in the NLO calculation and
eventually spoiling its convergence are $\log(p_T^2/m^2)$ terms.
They also need therefore to be resummed to all orders to allow for a
sensible phenomenological prediction. Such a resummation has been performed
along the following lines~\cite{cg}. 

One observes that in the large-$p_T$ limit ($p_T \gg m$) the
only important mass terms are those appearing in the logs, all the others being
power suppressed. This means that an alternative description of heavy quark
production can be achieved by considering {\sl massless} quarks and providing 
at the
same time perturbative distribution and fragmentation functions also for the 
heavy quark, describing the logarithmic mass dependence. The factorization
formula becomes
\beq
d\sigma(p_T) = \sum_{ijk} \int F_{i/H_1}(\mu,[m]) F_{j/H_2}(\mu,[m]) 
d\hat\sigma(ij\to k; p_T, \mu) D_k^Q(\mu,m),
\label{fact}
\eeq
with parton indices $i$,$j$ and $k$ also running on $Q$, taken massless in 
$\hat\sigma$, now an $\overline{\mathrm MS}$ 
subtracted cross section for light parton production.
The dependence on $m$ of the parton distribution functions $F_{i/H}$,  
shown between square brackets in eq. (\ref{fact}), is only there 
when $i$ or $j$ happens to be the heavy quark $Q$.

The key point is that the large mass of the heavy quark allows for 
the evaluation in perturbative QCD (pQCD) of its distribution and fragmentation 
functions.
Initial state conditions for $F_{Q/H}(\mu_0=m)$~\cite{ct}
and $D_k^Q(\mu_0\simeq m)$~\cite{melenason}
can be calculated in pQCD at NLO level in the $\overline{\rm MS}$ scheme. 
These initial state conditions can then be 
evolved with the Altarelli-Parisi  equations up to the large scale set by $\mu
\simeq p_T$. This evolution will resum to all orders the large logarithms
previously mentioned. 

\begin{figure}[t]
\begin{center}
\begin{minipage}{6.5truecm}
\epsfig{file= 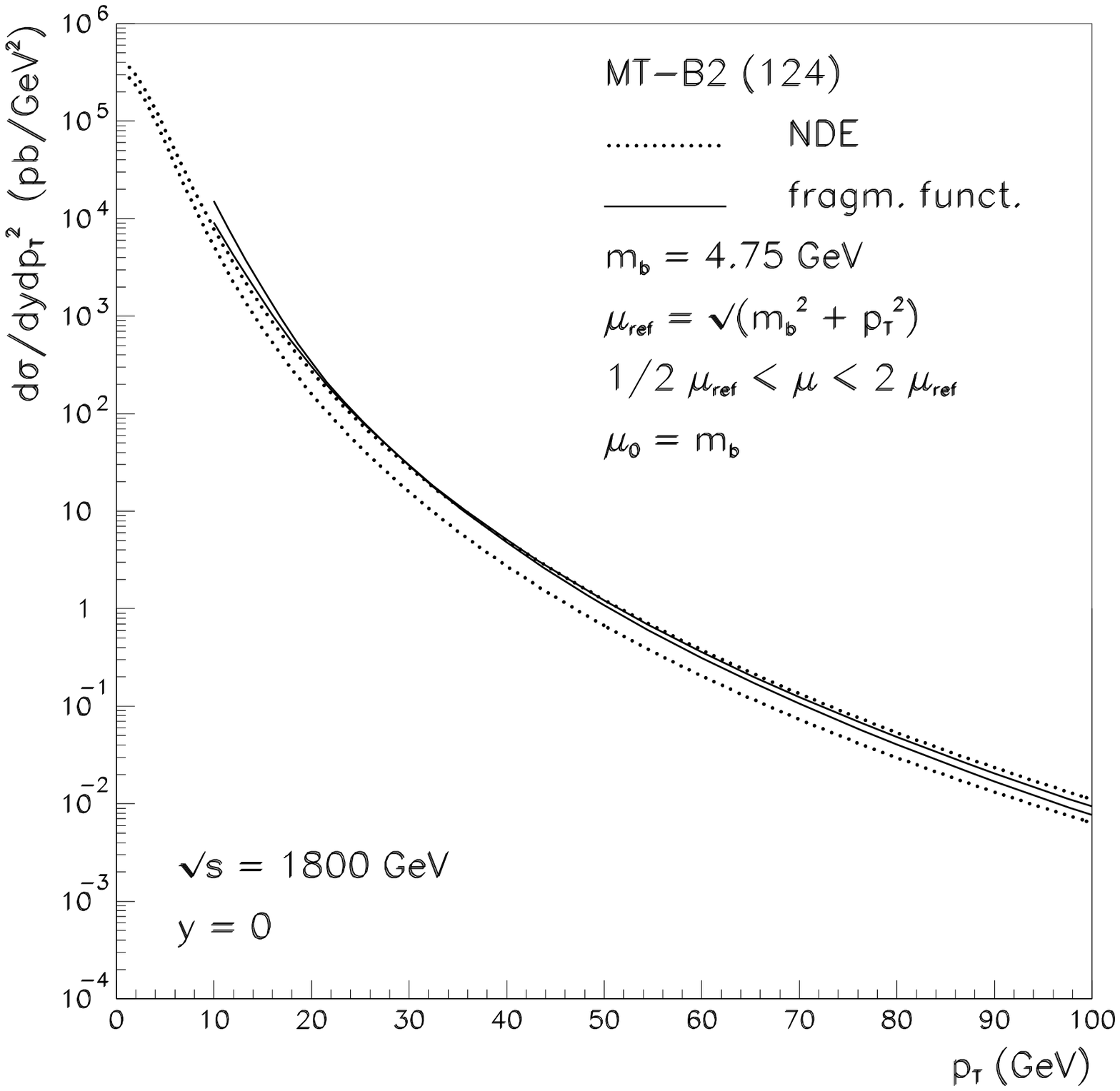,width=6.8truecm,height=5.7truecm,clip=}
\end{minipage}
\begin{minipage}{6.5truecm}
\epsfig{file=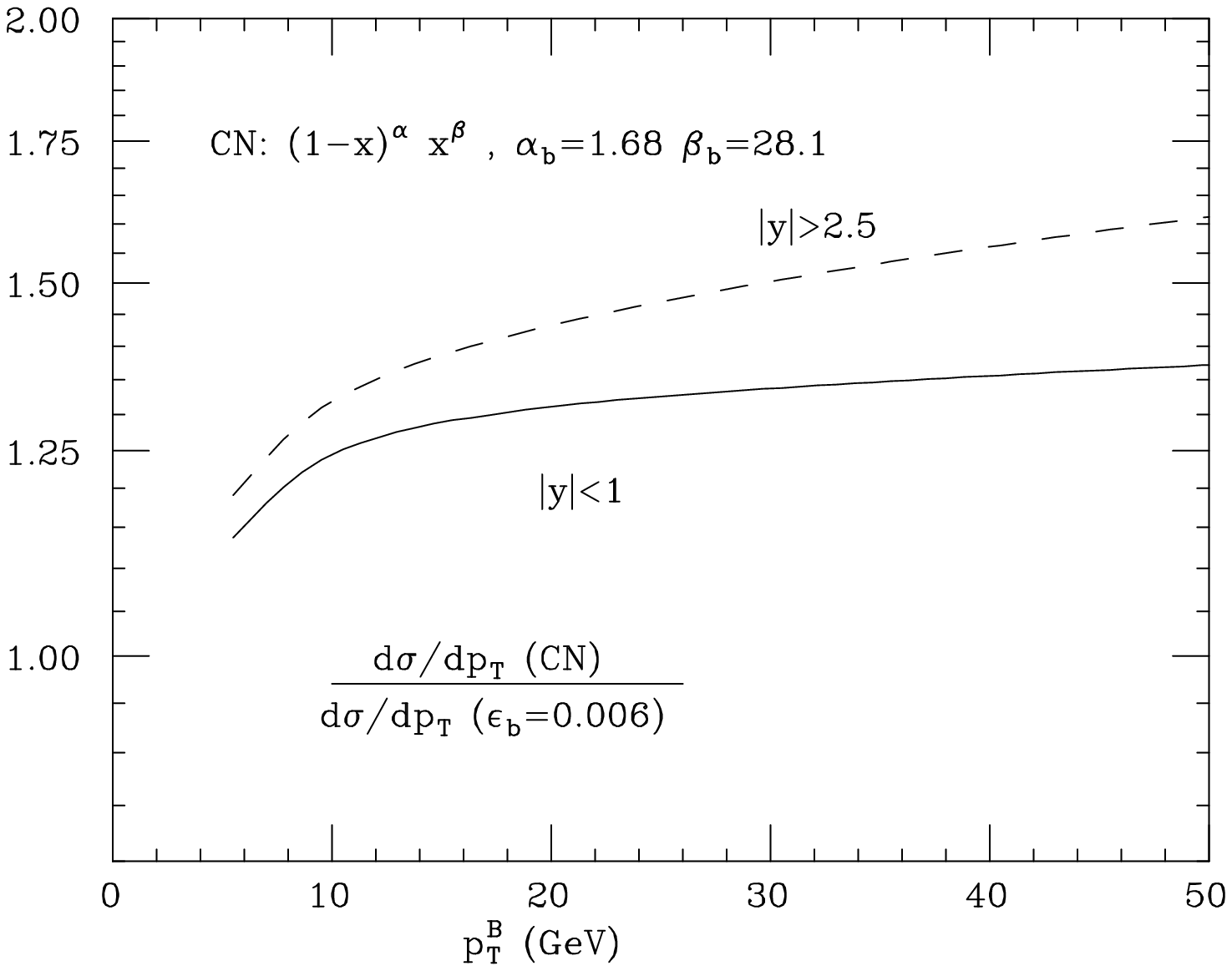,width=7.6truecm,height=5.6truecm,clip=}
\end{minipage}
\end{center}
\caption{Comparison of bottom $p_T$ distributions in the fixed order (NDE) and
resummed approaches (left plot, from ref. \protect\cite{cg}), and effect of a 
harder $b\to B$ fragmentation function on bottom production at the Tevatron
\protect\cite{mlm-priv} (right plot).}
\label{fig4}
\end{figure}

It is important to mention that due to the neglecting of power suppressed mass
terms this approach becomes unreliable when $p_T\simeq m$. In this region
only a case by case comparison with the full NLO massive calculation -- here
reliable and to be taken as a benchmark -- can tell
how accurate the resummed result is.

Phenomenological analyses \cite{cg} show that the effect of the resummation
becomes sizeable only at very large $p_T$, say greater than 50 GeV for bottom
hadroproduction at the Tevatron, resulting in a milder
factorization/renormalization scale dependence of the result and in a slightly
softer $p_T$ spectrum.

Non-perturbative effects describing, say, the $b\to B$ meson transition can
also be included within this formalism. The $b\to B$ fragmentation function can
be fitted to LEP $e^+e^-$ data and then used for predicting $B$ cross sections
at the Tevatron. Recent analyses \cite{cgee} show that this FF, when used in
connection with a NLO evaluation of bottom production like the MNR one, should
probably be taken harder than commonly done in the past. This choice of a
harder non-perturbative FF would increase the Tevatron cross sections from 
twenty to fifty per cent \cite{mlm-priv} (see fig.~\ref{fig4}), helping 
to reconcile theory and experimental data.

\section{Heavy quarkonium Production}

The production of heavy quarkonia has been subjected to intense study in the
last two or three years, with tens of papers having being
produced on the problem of $\jpsi$, $\psi'$, $\chi$ and $\Upsilon$ production in
$e^+e^-$, $\gamma p$, $p\bar p$, $pN$, $\pi N$ collisions.

The reason for such a surge in interest was the appearance of a theoretical
framework, the so called Factorization Approach (FA) by Bodwin, Braaten and
Lepage \cite{bbl}, which seems able to solve  the theoretical problems
that quarkonium production models faced in the past, and also to reconcile
theoretical predictions with experimental data, previously in disagreement up
to factors of fifty in some instances.

In this talk I shall not review the Factorization Approach in detail, leaving
this theoretical introduction to other sources (see for instance 
\cite{onium}).
%
I shall just recall how the FA
writes the quarkonium state $H$ production cross section in the following form:
\beq
\sigma(ij\to Q\overline{Q}\to H) = \sum_n \hat\sigma(ij\to Q\bar
Q[n])\langle{\cal O}^H(n)\rangle .
\eeq
According to this equation, the cross section for producing the observable
quarkonium state $H$ is factorized into two parts. In the short distance part a
$Q\overline{Q}$ pair of heavy quarks is produced in the spin/colour/angular momentum
state $^{2S+1}L_J^{(c)} \equiv n$ by the scattering of the two light partons
$i$ and $j$. Subsequently this pair hadronizes into the quarkonium $H$;
$\langle{\cal O}^H(n)\rangle$ is formally a non-relativistic QCD (NRQCD) matrix
element describing this non-perturbative transition.

An important feature of this equation is that also $Q\overline{Q}$ pairs in a
colour octet state are  allowed to contribute to the production of a colour
singlet quarkonium $H$: their colour is neutralized via a non-perturbative
emission of soft gluons. While the corresponding matrix elements are suppressed
by the need of such an emission, the short distance coefficients can on the
other hand be large, perhaps overcompensating the suppression of the 
non-perturbative term. This explains why colour octet contributions can play a very
important role in predicting the total size of quarkonium production cross
sections.

The limited space available doesn't allow  any detail about the
phenomenological studies which have been performed. Ref. \cite{moriond}
contains a small up-to-date review, with references to the original papers. 
I shall only mention here that the introduction
of the colour octet channels allows for what looks like a ``reasonable''
description of the data. Colour octet matrix elements have to be fitted to the
data themselves, and the uncertainties on these fits are certainly not smaller
than a factor of two, due to the many systematics entering their determination:
parton distribution functions,  charm quark mass, $\alpha_s$ value, higher
order QCD corrections, etc. However, the values one gets appear of the correct
order of magnitude if compared via NRQCD scaling rules \cite{LMNMH92} to the 
colour singlet ones, thereby supporting the underlying picture.

So far all fits have been performed using leading order cross sections. 
Very recently a next-to-leading order calculation for quarkonium total cross
sections, both via singlet and octet channels, has been completed \cite{pcgmm}.
It will therefore be possible to reduce at least some of the uncertainties
previously mentioned and hence obtain more reliable fits.

\end{document}